
%
%
\magnification=\magstep1
\advance\voffset15mm
\advance\mathsurround1pt
%
%
\newskip\diaskip \diaskip=9mm
\def\gev{\hbox{\rm  GeV}}
\def\gmn{g_{\mu\nu}}
\def\into{I_0}
\def\subint#1{c^{\phantom2}_{#1}}
\def\mt{m^2}
\def\pt{p^2}
\def\qt{q^2}
\def\pmt{(p^2-m^2)}
\def\intdp{\int\!d^4p\,\,\,}
\def\dpi{{1\over(2\pi)^4}\delta^4}
\def\pimm{{\Pi}_\mu^{\!\phantom{\mu}\mu}(0)}
\def\dmud{\partial_\mu}

\def\dmuu{\partial^\mu}

\def\drho{\partial_\rho}
\def\dl{\partial_\l}
\let\phi\varphi
\def\psibar{\bar\psi}
\let\cl\centerline
\let\eps\epsilon
\def\mw{\hbox{$m_{\rm W}^{\phantom2}$}}
\def\mtop{\hbox{$m_{\rm t}^{\phantom2}$}}
\def\mwsq{\hbox{$m_{\rm W}^{\phantom2}{}^{2}$}}
\def\mtopsq{\hbox{$m_{\rm t}^{\,\,2}$}}
\def\ref#1{\hbox{${}^{#1}$)}}
\def\Tr{\mathop{\hbox{\rm Tr}}}
\def\mpht{m_\gamma{}^{\!2\,}}
\def\phat{\hbox{$\not\mkern-3.3mu p\mkern3.3mu$}}
\def\L{{\cal L}}
\let\l\lambda
\let\s\sigma
\def\tw{\theta_W^{\phantom2}}
\def\Feynman{\hfill {Feynman diagram to be added}\quad}

\font\ninerm=cmr9
\headline={\it J. J. Lodder, Preprint. \hfill
hep-th/9405265
\hfill
Submitted to Physica\thinspace {\bf A}, 1994, January 6}

\cl{\bf\uppercase{Quantum-Electrodynamics without
renormalization}}
\medskip
\cl{\bf\uppercase{V. Bosonic contributions
to the photon mass}}
\bigskip
\cl{ J. J. LODDER}
\smallskip
\cl{\it Oudegracht 331\thinspace bis,
3511\thinspace\thinspace PC Utrecht,
The Netherlands.}
\bigskip\bigskip\bigskip
{\ninerm
The bosonic contributions to the photon mass are
shown to be
of the same form as the fermionic
contributions,
but of opposite sign.
A mass sum rule
for bosons and fermions follows from gauge invariance.
Assuming completeness of the standard model
the top quark mass can be predicted.
Effectively only the W-bosons contribute,
giving a lowest order prediction
of~$85.1$~$\gev$
for the mass of top quark,
on the low side of currently accepted estimates.
}

\bigskip
\noindent{\bf 1. Introduction}
\bigskip
In the preceding parts (I--IV) of this series\ref{2} it
was shown that the symmetrical theory of generalised
functions\ref{1,3} can be applied to computations
in quantum field theory.
This paper can be read without consulting the
preceding parts of the series.
The needed results are summarized briefly.

The symmetrical theory of generalised functions never has
infinity in any result,
and the available simple model\ref3 is strong enough to
handle all special functions which occur in quantum field
theory,
so divergencies do not occur when this
theory is applied to quantum field theory. All results are
automatically finite and renormalization is never needed
to remove infinities. Infinity of integrals is replaced by
the less restricted concept of determinacy\ref3, which is
determined by the scale transformation properties.

In part III of this series the
photon mass was found to be determinate, and therefore
physically relevant. It is (of course) finite and it is
found to be non-zero, in disagreement with gauge
invariance and therefore with all experience.

An attempt (in part III) to obtain cancellation between
orders,
and thereby an equation fixing the fine-structure
constant~$\alpha$, did not lead to useful results.
In this paper the contribution of charged bosons to the
photon mass is evaluated
and it is shown that bosonic contributions may cancel the
fermionic contributions,
provided that a mass sum rule holds.
If this idea is realized in Nature it is possible
to predict the mass of the top quark.
The result is~$\mtop=85.1\pm0.3\,\,\gev$ to leading order.

\bigskip
\noindent{\bf 2. Fermions}
\bigskip
In the previous parts of this series the
convention~$\gmn=(-,+,+,+)$ was used,
in agreement with most
of the original QED literature.
In this paper the computations are transcribed to the
nowadays more popular convention~\hbox{$\gmn=(+,-,-,-)$}.
\goodbreak
\noindent
The fermionic
contribution to the photon mass~$m_\gamma$
is found from
\vskip\diaskip
\vbox to 0pt{\vss\line{\Feynman \hfill(1)}}
\vskip\diaskip
\noindent
with photon momentum~$k=0$.
Substitution of the Feynman rules for Dirac fermions from Ap.\thinspace A
({\it including the fermion loop minus sign}\/) gives
$$i\,\mpht\gmn=\Pi_{\mu\nu}(0)=
-4\pi\alpha q^2\Tr\int\!{d^4p\over(2\pi)^4}
\,\,\,\gamma_\mu
{\phat+m\over\pmt}\gamma_\nu{\phat+m\over\pmt},
\eqno{(2)}$$
in which~$\alpha$ is the fine-structure constant,
and~$q^2$ is the squared relative charge of the fermion.
It is convenient to contract with~$g^{\mu\nu}$ to
obtain
$$4i\,\mpht=\pimm=2{\alpha q^2\over\pi^3}\mt
\intdp{p^2-2m^2\over\mt(p^2-m^2)^2}:=
2\alpha \qt\mt\into/\pi^3,
\eqno{(3)}$$
where~$\into$ is a convenient abbreviation for the
remaining dimensionless integral.
$$\into:=
\intdp{p^2-2m^2\over\mt(p^2-m^2)^2}=
-i\,\pi^2,\eqno{(4)}$$
which was shown to be finite,
non-zero,
and determinate in part III,
despite its quadratic divergence.
Correcting for the new normalization
the integral equals~$+i\pi^2$ by (I-E.5,11),
so the squared photon mass correction produced by
fermions
of mass~$m$ is
$$\mpht={\alpha q^2\over2\pi}\mt,\eqno{(5)}$$
as derived~(III.7) before.
This result was first obtained and printed
(as far as I am aware)
by Keller\ref4.
It seems likely
that this result has been found before
without being published,
since (as commonly understood)
it is in complete disagreement with all experience.

\bigskip
\noindent{\bf 3. Scalar bosons}
\bigskip
The Lagrangian and the Feynman rules are summarized in
appendix A. Instead of the simple diagram~(1) for fermions
there are now two contributions to the photon self-energy,
the analogue of the fermion  loop, and the bubble diagram
\vskip\diaskip
\vbox to 0pt{\vss\line{\Feynman\hfill(6)}}
\vskip\diaskip
\noindent
Putting~$k=0$ and contracting with~$g^{\mu\nu}$
gives the integral
$$\pimm={\alpha q^2\over\pi^3}
\intdp{\pt\over\pmt^2},\eqno{(7)}$$
for the loop diagram and
$$\pimm=-2{\alpha q^2\over\pi^3}\intdp{1\over
(p^2-m^2)},\eqno{(8)}$$
for the bubble diagram. Combining these gives the total
contribution
$$\pimm={}-{\alpha q^2\over\pi^3}\mt\!
\intdp{\pt-2\mt\over\mt\pmt^2}=
{}-\alpha q^2m^2\into/\pi^3.\eqno{(9)}$$

Comparing this result with the fermion contribution~(3)
it is seen that the contribution of charged scalar bosons
to the photon mass equals {\it minus}\/ half the contribution from a
spin~${1\over2}$ fermion of the same mass.

The minus
sign is due to the fact that fermion loops carry an
additional minus sign, which boson loops do not have.
The scalar boson contribution is determinate and finite,
as it is
for fermions,
since it is proportional to the same integral.

\bigskip
\noindent{\bf 4. Vector bosons}
\bigskip
Fundamental charged scalar bosons
are not known,
but charged intermediate vector bosons
are seen as carriers of the
weak interaction.
For vector bosons there are again two Feynman diagrams
contributing to the
photon self energy
\vskip\diaskip
\vbox to 0pt{\vss\line{\Feynman\hfill(10)}}
\vskip\diaskip
\noindent
obtained from the vertices~({\bf A}12,13).
It is not clear in advance
that the vector boson contribution will be
proportional to the same integral,
since the first diagram contains terms with (in power
counting terminology)
logarithmic, quadratic, quartic, and even sextic
degree of divergence,
while the bubble diagram has quartic, quadratic and
logarithmic terms.
However, in the first diagram the sextic terms
involving~$p^4p_\mu p_\nu $ cancel, leaving upon contraction
with~$\gmn$ the quartic and quadratic terms
$$\pimm=
{}-3{\alpha q^2\over\pi^3}
\intdp{p^4-3\pt\mt\over2\mt\pmt^2},\eqno{(11)}$$
with the W-subscript on the mass omitted for clarity.
The seagull vertex~({\bf A}13) gives the bubble diagram,
which yields upon substitution
$$\pimm=3{\alpha q^2\over\pi^3}
\intdp{\pt-4\mt\over2\mt(\pt-\mt)}.\eqno{(12)}$$
The quartic terms cancel in the sum of the two
diagrams,
so the total vector boson
contribution to the photon
mass is
$$\pimm={}-3{\alpha q^2\over\pi^3}
\mt\!\intdp{\pt-2\mt\over\mt\pmt^2}=
{}-3\alpha\qt\mt\into/\pi^3,\eqno{(13)}$$
which is again proportional to the same
determinate quadratic integral~(3) found in the
fermionic case.
It may be noted that the freedom from subtractions
and the full linearity
of the symmetrical theory of generalised functions
makes it unnecessary to invent
special tricks
to avoid the canceling terms in~(11) and~(12).
\bigskip
\noindent{\bf 5. The mass sum rule}
\bigskip
Since the boson
and fermion contributions to the photon mass have opposite
sign they will cancel if the mass sum rule
$$\sum_{\hbox to 0pt{\sevenrm\hss fermions\hss}}
\,g_f^{\phantom{2}}q_{\!f}^{\,2}\subint jm_{\!f}^{\,2}=
\sum_{\hbox to 0pt{\sevenrm \hss bosons\hss}}
\,g_{b}^{\phantom{2}}q_b^{\,2}\subint jm_b^{\,2},
\eqno{(14)}$$
is satisfied.
The sums are over all fundamental fermions and bosons.
The factors~$g_f$ and~$g_b$ are the multiplicity of the
particles, and the~$q^2$'s are the charges measured in units of
the squared electron charge.
The factors~$c_j$ are the relative coefficients
of the integrals obtained from
the vacuum polarization diagrams,
which were calculated~(3,13) in the previous sections.

If the leptons, quarks and the W-bosons
of the standard model
are assumed to be the only fundamental charged particles,
all masses in~(14) are known with the exception
of the top quark mass.
The top mass can therefore be calculated from the sum rule.

To presently available accuracy all other quarks and
leptons are effectively (squared) massless,
leaving only the top- and the W-contribution.
Taking the~$W^\pm$ bosons to be a unit charge doublet
we must nevertheless take~$g_W^{\phantom2}{=}1$.
The required factor~$2$ has already been accounted
for in the Feynman rules~({\bf A}12,13).
Assuming threefold colour degeneracy and charge 2/3 for
the top quark
substitution of the integral
coefficients~$\subint {\!1/2}{=}2$,~$\subint1{=}3$ yields
$$\mtopsq = {9\over8} \mwsq,\eqno{(15)}$$
or
$$\mtop={3\over4}\sqrt2\,\,\mw.\eqno{(16)}$$
Substitution the value\ref5~$\mw=80.22\pm0.26~\gev$
for the W-mass yields a top mass
of
$$\mtop=85.1\pm0.3\,\,\gev
\quad{}+\,\,\cdots\cdots.\eqno{(17)}$$
The given error reflects the experimental
W-mass uncertainty only.
In addition there will be higher order
electro-weak and strong corrections
which have not yet
been estimated,
which will probably be larger than the experimental
W-mass uncertainty.

The result is on the low side of currently
accepted estimates\ref5 of the top mass,
but these estimates
are based on difficult experiments,
which in addition
depend on difficult theoretical computations for
their interpretation,
so a judgement on the correctness of
the prediction~(17) must await experiments which can
observe the top quark unambigeously by obtaining a
measurement of its mass.
This is expected to be possible in the near future.
\goodbreak

\bigskip\noindent
{\bf 6. Discussion}

The photon mass has been a source of difficulties
in QED from
the very beginning.
It is obvious upon computation
that the photon mass is not zero
automatically,
as required by charge conservation and gauge invariance,
so it needs special treatment.

Special renormalization methods have been devised to get rid
of it, as discussed in part~IV of this series
of papers.
The best known methods,
in order of appearance,
are brute force subtraction,
Pauli-Villars regularization,
or dimensional regularization.
These methods all lack a natural mathematical
and physical interpretation.

For the argument presented in this paper the method of
regularization is irrelevant,
since all contributions to the photon mass are
proportional to the same integral~(4).
For any non-zero interpretation of this integral the mass
sum rule~(14) leads to a vanishing photon mass,
and conversely the sum rule
makes a forced evaluation to zero
unnecessary.

The determinacy of the integral~(4)
in the generalised function framework
makes it impossible to put it
equal to zero,
so the validity of the sum rule~(14)
is the only way left to obtain gauge invariance.
A proof of determinacy to all orders and of the
sufficiency of~(14) in higher orders remains to be given,
but the result can be made plausible\ref2.

Sum rules such as~(14) cannot be falsified
until we possess a complete theory of everything.
Conversely
experimental verification
of the sum rule would imply that the standard
model is probably complete.
Additional bosons and fermions would again have to satisfy
the sum rule by themselves,
which seems unlikely.

If the top quark is not found
at the predicted mass,
heavier charged bosons are necessary
to balance the sum rule.
Charged Higgs bosons are a possibility
which can be accommodated in the standard model.
If this happens to be realized
it will take much longer
until the relevant experiments can be performed.

The derivation given in this paper is clearly inadequate,
since the validity of
a fundamental result like the mass sum rule (14)
should be clear beforehand,
instead of appearing afterwards in perturbation theory.
This will have to await the development of
a more fundamental theory from which the mass spectrum
of the quarks, leptons, and vector bosons can be derived.
Nevertheless, the electromagnetic coupling of the charges
should be given correctly
even by the present incomplete theory,
so the mass sum rule~(14) may be expected to hold also
in future,
more complete theories.
(Application of minimal coupling to the generic
Proca Lagrangian
yields the same electromagnetic interaction).

It is interesting to see that meaningful cancellation of
bosonic and fermionic contributions to undesirable results
may occur without invoking supersymmetry.
There is a difference;
with supersymmetry one hopes
for cancellation of undesirable infinities,
in the symmetrical theory of generalised functions
applied to quantum field theory
there are no infinities.
Instead one has cancellation
of unobserved physical predictions.

It remains to be seen if the value of the top quark mass
of~$85.1\gev\!$,
predicted from gauge invariance
and the assumed completeness of the standard model,
will be confirmed
by experiment.
\bigskip
\noindent{\bf Acknowledgement}
\medskip\noindent
Dr.~H. J. de Blank is thanked
for critical discussion and improvement of the manuscript.

\goodbreak

\bigskip
\noindent
{\bf Appendix A\ \  Feynman rules}
\bigskip
For ease of reference the Feynman rules
for charged fermions and
bosons with the modern choice~$\gmn=(1,-1,-1,-1)$ for the
metric tensor.

\noindent
For charged scalar bosons with Lagrangian\ref6
$$\L=\dmud\phi^\dagger\,\dmuu\phi-
\mt\phi^\dagger\phi,\eqno{({\bf A}1)}$$
the assumption of minimal coupling leads to the
electromagnetic interaction Lagrangian
$$\L_{\rm int}=ieA^\mu(\dmud\phi^\dagger\,\phi
-\phi^\dagger\,\dmud\phi)
+e^2A^2\phi^\dagger\phi.\eqno{({\bf A}2)}$$
The standard derivation of the Feynman rules gives the
wave equation and propagator
$$(\dmuu\dmud+\mt)\phi=0,\quad\hbox{and}\quad
{i\over\pmt},\eqno{({\bf A}3)}$$
the three wave vertex factor
\vskip\diaskip
\vbox to 0pt{\vss\line{\hfill{\rm 3-Vertex}$\displaystyle
{}=i(p_{1\mu}-p_{2\mu})\dpi(k+p_1+p_2),$\hfill({\bf A}4)}}
\vskip\diaskip
\noindent
and the four wave (or seagull) vertex factor
\vskip\diaskip
\vbox to 0pt{\vss\line{\hfill\rm 4-Vertex$\displaystyle
=2ie^2\gmn\dpi(k_1+k_2+p_1+p_2),
$\hfill({\bf A}5)}}
\vskip\diaskip
\noindent
with all momenta taken as incoming.
\smallskip\noindent
For charged Dirac fermions with free Lagrangian
and minimal electromagnetic coupling
$$\L=i\psibar\gamma_\mu\dmuu\psi-m\psibar\psi
-e\psibar\gamma_\mu\psi A^\mu,\eqno{({\bf A}6)}$$
the standard derivation of the Feynman rules\ref6 gives the
Dirac equation and propagator
$$i\gamma_\mu\dmuu\psi-m\psi=0
\qquad\hbox{and}\qquad
i{\phat +m\over\pmt},\eqno{({\bf A}7)}$$
and the electromagnetic fermion vertex factor
\vskip\diaskip
\vbox to 0pt{\vss\line{\hfill\rm 3-Vertex $\displaystyle
{}=i\,{\rm e}\,\gamma^\mu\dpi(k+p_1+p_2).$\hfill({\bf A}8)}}
\vskip\diaskip
\noindent
There is no four wave coupling for fermions.

\smallskip\noindent
For intermediate vector bosons the relevant part of
the Lagrangian\ref7 is
$$\L={1\over2}\Tr\bigl( W_{\l\rho}W^{\l\rho}\bigr)
+\mwsq W^+_\l W^{-\,\l}.
\eqno{({\bf A}9)}$$
The field strength tensor is given by
$$
W_{\l}=W_{\l}^a\tau_a/2,\qquad
W_{\l\rho}^a=\dl W_\rho^a-\drho W_\l^a -g\eps_{abc}
W_\l^bW_\rho^c,\eqno{({\bf A}10)}$$
with~$g$ the weak coupling constant,
$\tau_a$ the Pauli matrices,
and~$\eps_{abc}$ the completely antisymmetric SU(2)
structure constants.
Working out
the product and computing
the Feynman rules (in the unitary gauge)
gives the propagator
\vskip\diaskip
\vbox to 0pt{\vss\line{\hfill\rm3-Vertex$\displaystyle
{}={-i\over \mwsq}\,\,
{\mwsq\gmn-p_\mu p_\nu\over \pt-\mwsq},$\hfill({\bf A}11)}}
\vskip\diaskip
\noindent
and with~$W_\l^\pm=(W_\l^1\mp iW_\l^2)/\sqrt2$,
$W_\l^3=A_\l\sin\tw$,
$g\sin\tw=e$,
the three boson vertex coupling the~$W^\pm$
to the photon becomes
\vskip\diaskip
\vbox to 0pt{\vss\line{\hfill\rm3-Vertex$\displaystyle
{}=ie\big((k-p_1)_\sigma g_{\mu\nu}
+(p_1-p_2)_\mu g_{\rho\sigma}
+(p_2-k)_\rho g_{\mu\sigma}\big)
$\hfill({\bf A}12)}}
\vskip\diaskip
\noindent
with the momentum conserving~$\delta$-function understood.
The quadratic term in~$g$ gives the four boson vertex
coupling the ~$W^\pm$ to two photons
\vskip\diaskip
\vbox to 0pt{\vss\line{\hfill\rm4-Vertex$\displaystyle
{}=-ie^2(2g_{\mu\nu}g_{\rho\s}
-g_{\mu\rho}g_{\nu\s}
-g_{\mu\s}g_{\nu\rho}),
$\hfill({\bf A}13)}}
\vskip\diaskip
\noindent
which are analogous to the corresponding terms in the
scalar case.
The boson vertex factors depend on the momenta
at the vertex, giving rise to stronger divergence.
\bigskip
\goodbreak
\noindent{\bf References}
\medskip

\parindent=-1.5em
\leftskip=1.5em
\def\ref#1{\par\leavevmode\rlap{#1)}\kern1.5em}
\def\aut#1{{\rm#1}}                     \let\author        =\aut
\def\tit#1{{\sl#1}}                     \let\title          =\tit
\def\yea#1{{\rm(#1)}}   \let\year            =\yea
\def\pub#1{{\rm#1}}                     
\def\jou#1{{\rm#1}}                     
\def\vol#1{{\bf#1}}                     
\def\pag#1{{\rm#1}}                     \let\page            =\pag

\let\lin\relax

\ref{1}
 \aut{Lodder J.J.},
 \lin\jou{Physica}
 \vol{116A}
 \yea{1982} \pag{45, 59, 380, 392},
 \jou{Physica}
 \vol{132A}
 \yea{1985} \pag{318}.

\ref{2}
 \aut{Lodder J.J.},
 \lin\jou{Physica}
 \vol{120A}
 \yea{1983} \pag{1, 30, 566, 579}.

\ref{3}
 \aut{Lodder J.J.},
 \tit{Towards a Symmetrical Theory
     of Generalised Functions},
 \jou{CWI tract }\vol{79}
 \yea{1991},
 \pub{CWI, Amsterdam}.

\ref4
 \aut{Keller, K.},
 \tit{Konstruction von Produkten in einer f\"ur
    Feldtheorien wichtige Klasse von Distributionen},
 \pub{Thesis, Aachen (unpublished)},
 \year{1974}.

\ref5
 \aut{Particle Data Group},
 \jou{Phys. Rev.}
 \vol{D45}
 \yea{1992}
 \page{part 2.}

\ref{6}\aut{Itzykson  C., and Zuber J.B.},
 \title{Quantum Field Theory},
 \pub{McGraw Hill, New York} \year{1980}.

\ref{7}
\aut{Nachtmann, O.},
 \tit{Elementary Particle Physics},
 \pub{Springer Verlag, Berlin},
\yea{1990}.
\eject

\medskip\noindent
\bf{Note added 10-5-94}
\smallskip\noindent
If the top quark is heavier than~$85.1$ \gev\
it must be much heavier,
since light charged bosons are known not to exist.
Assuming charged scalar bosons the missing mass is
$$m^2_H={8\over3}\mtopsq-3\mwsq,$$
which yields~$m_H=205$
\gev for ~$170$ \gev top quarks.
Charged Higgs bosons may be a possibility.

\bye